\documentclass[CRPHYS,Unicode,manuscript]{cedram}

\usepackage{subcaption}
\usepackage{tikz}
\usepackage{bm}
\usetikzlibrary{shapes.geometric, arrows}

\newcommand\munu{{\mu\nu}}

\newcommand\fracd[2]{\frac{d {#1}}{d {#2}}}

\newcommand\cald{\mathcal D}
\newcommand\calo{\mathcal O}

\newcommand\tensor[1]{\boldsymbol{#1}}
\newcommand\eye{\tensor 1}

\newcommand\uvec[1]{\hat {#1}}

\newsavebox{\mySaveBoxMath} 
\newsavebox{\mySaveBoxText} 

\renewcommand{\vec}[1]{\boldsymbol{#1}}
\renewcommand{\uvec}[1]{\hat{\boldsymbol{#1}}}
\renewcommand{\tensor}[1]{\underline{\boldsymbol{#1}}}

\newcommand{\compactcross}[1]{\hspace[-0.1em]\times\hspace[-0.1em]}

\title{Ray tracing methods for wave propagation in moving anisotropic media : application to magnetized plasmas}

\alttitle{Méthodes de lancer de rayon pour la propagation des ondes dans les milieux anisotropes en mouvement : application aux plasmas magnétisés}

\author{\firstname{Aymeric} \lastname{Braud}\IsCorresp\CDRorcid{0009-0008-8802-7393}}
\address{LAPLACE, Université de Toulouse, CNRS, INPT, UPS, 31062 Toulouse, France}
\thanks{This work is supported by the Agence Nationale de la Recherche (ANR) through the WaRP project (ANR-21-CE30-0002).}
\email[]{aymeric.braud@laplace.univ-tlse.fr}
\CDRGrant[ANR]{ANR-21-CE30-0002}

\author{\firstname{Julien} \lastname{Langlois}\CDRorcid{0009-0008-8387-9247}}
\addressSameAs{1}{LAPLACE, Université de Toulouse, CNRS, INPT, UPS, 31062 Toulouse, France}

\author{\firstname{Renaud} \lastname{Gueroult}\CDRorcid{0000-0001-5208-9594}}
\addressSameAs{1}{LAPLACE, Université de Toulouse, CNRS, INPT, UPS, 31062 Toulouse, France}

\altkeywords{Optique géométrique, entraînement de la lumière, écoulement plasma, diélectrique en mouvement}

\begin{abstract}
  The propagation of a wave in a medium is generally affected when the medium is moving with respect to the observer. Because plasma equilibria often involve plasma flows, for instance in astrophysics or in magnetic confinement nuclear fusion devices, understanding the effect of motion on plasma waves is important. Meanwhile, the presence of a background magnetic field in a plasma makes it anisotropic. To address this problem, we derive here ray tracing equations for the trajectory of rays propagating in a moving anisotropic medium. The proposed approach is to use an effective dispersion relation for the moving medium as seen from the laboratory, obtained by performing a Lorentz transformation of the dispersion relation known for the medium at rest. This formalism is illustrated by considering the standard ordinary and extraordinary modes in a magnetized plasma at rest. Although we work here at lowest order in the geometrical optics approximation, this method is a first step towards higher order expansions, as required for instance to capture polarization effects.
\end{abstract}

\begin{altabstract}
  La propagation d'une onde dans un milieu est en général modifiée lorsque celui-ci est en mouvement. Les configurations d'équilibre d'un plasma reposant souvent sur un champ de vitesse, comme par exemple en astrophysique ou pour la fusion par confinement magnétique, une compréhension des effets du mouvement sur les ondes plasmas est particulièrement souhaitable. On s'intéresse ici à développer une méthode de lancer de rayon pour étudier la trajectoire des rayons se propageant dans un milieu anisotrope en mouvement dans l'approximation de l'optique géométrique. Une relation de dispersion effective pour le milieu en mouvement vu du laboratoire est identifiée en effectuant une transformation de Lorentz de la relation de dispersion du milieu au repos. Cette relation de dispersion est alors utilisée dans des équations de lancer de rayon, permettant ainsi de modéliser l'effet du mouvement sur la trajectoire des différents modes supportés par le milieu. Le potentiel de cette méthode est pour finir illustré en considérant les modes ordinaire et extraordinaire classiques d'un plasma magnétisé. Ces travaux, même si limités à l'ordre le plus bas de l'optique géométrique, offrent les premiers élements d'une théorie étendue aux ordres supérieurs, qui permettra alors de capturer entre autres choses les effets sur la polarisation.
\end{altabstract}

\begin{document}

\maketitle

\section{Context and motivations}

Waves are used extensively in plasmas. Important applications of plasma waves include plasma control, for instance for plasma heating and current drive in magnetic confinement fusion devices~\cite{Fisch1987}, as well as plasma diagnostics, such as estimating interstellar magnetic fields via Faraday rotation~\cite{Lyne1968,Han2018}. Waves may also offer new means to drive plasma rotation~\cite{Ochs2021,Ochs2021a,Ochs2022,Rax2023,Ochs2024}. The design of these control systems and the interpretation of these diagnostics rely directly on plasma waves theory to model propagation in these varied environments. Since plasmas are anisotropic when immersed in a background magnetic field, these propagation models classically build on the theory of waves in anisotropic dispersive media, in the presence of possible plasma non-uniformities (density, magnetic field, etc.). 

 Other than for rare exceptions, these models and the theory of waves in plasmas however neglect the effect of a velocity field. Indeed, even though it has long been established that motion can have an effect on wave propagation~\cite{fresnel_lettre_1818,fizeau_sur_1851,jones_aether_1975}, and that rotation phenomena are encountered and play a key role in a wide range of environments from laboratory plasmas to astrophysics~\cite{Miesch2009} to magnetic confinement fusion~\cite{Strait1995}, wave propagation properties are for the most part determined assuming a plasma at rest. Examining this apparent shortcoming, it has been confirmed in recent years that motion and in particular rotation can lead to a number of peculiar wave manifestations in plasmas~\cite{gueroult_wave_2023}, including effects on the wave transverse structure~\cite{rax_faraday-fresnel_2021,Rax2023b} and polarization~\cite{gueroult_determining_2019,gueroult_enhanced_2020}, and modifications to wave-particle resonance conditions~\cite{Rax2023a}. Although promising, these analytical studies are limited to simple configurations (e.~g. geometry and plasma response). The aim of the work presented here is to present the first elements of an eikonal formalism 
for waves in a moving anisotropic medium, which could eventually allow simulating wave propagation in moving plasmas in more complex configurations.

This manuscript is organized as follows. First, in Section~\ref{Sec:SecII}, we recall the basic elements of geometrical optics for static media, pointing how the ray tracing equations (RTEs) for the trajectory can be derived from the dispersion function obtained from a zeroth-order eikonal development. Then, in Section~\ref{Sec:SecIII}, we show how a moving medium can be modelled through a medium that is at rest in the lab-frame but bestowed with additional motion-dependent properties, and how the dispersion function for this effective medium can be obtained using covariance properties. Putting these pieces together, Section~\ref{Sec:SecIV} derives the trajectory RTEs for a moving medium, showing how classical results are recovered in the limit of an isotropic medium. Finally, Section~\ref{Sec:SecV} applies these findings to the case of a magnetized plasma. Lastly, some concluding remarks are given in Section~\ref{sec:4_conclusion}.


\section{Basics of geometrical optics}
\label{Sec:SecII}

As an introduction to the discussion in the next sections of the generalization of ray tracing equations to moving media, we first recall here some of the key elements of geometrical optics theory. This discussion follows primarily Tracy's textbook~\cite{tracy_ray_2014}, and interested readers are referred to Refs.~\cite{tracy_ray_2014,littlejohn_geometric_1991,rozanov_geometrical_2005} for an in-depth presentation of these concepts.

\subsection{Problem statement}

The physics problem of interest here is the propagation of waves in a linear medium. The multicomponent wave field is denoted $\vec \Psi$ and can be for example the electric field. Let us assume that this problem is entirely described by a linear wave equation\footnote{Rigorously, the general linear wave equation is an integrodifferential equation but can be reduced to the pseudodifferential form \eqref{Eq:wave_eq} using the Weyl symbol calculus. Moreover, for the sake of simplicity, we assume here that it does not depend on time $t$ although the method would be the same. Interested readers is referred to Ref.~\cite{tracy_ray_2014} for a fuller discussion of these two points.}~\cite{tracy_ray_2014}
\begin{equation}
\widetilde{\tensor{D}}(\vec x,i\vec\nabla,-i\partial_t)\vec\Psi=\vec 0
\label{Eq:wave_eq}
\end{equation}
with $\widetilde{\tensor{D}}$ a matrix of differential operators that can depend on space. The partial differential equation (PDE) Eq.~\eqref{Eq:wave_eq} describes how waves propagate in a given medium. 

To make it less abstract, we consider as an illustration a homogeneous isotropic non-dispersive medium with refractive index $n$. The wave equation for the electric field $\vec E$ in this medium is then 
\begin{equation}
\vec\nabla\times\vec\nabla\times\vec E+\frac{n^2}{c^2}\partial^2_t\vec E=0,
\end{equation}
so that in this case 
\begin{equation}
\widetilde{\tensor{D}}=\tensor\nabla\tensor\nabla+\frac{n^2}{c^2}\partial^2_t\eye
\end{equation}
with $\tensor\nabla$ the skew-symmetric matrix associated to the curl operator.

\subsection{Eikonal expansion and geometrical optics}

Geometrical optics (GO) is a method to obtain approximate solutions to the wave equation. The central characteristic of geometrical optics is the assumption of a scale separation, namely that the characteristic variation length $L$ of the properties of the medium is large compared to the wavelength $\lambda$ of the wave (i.~e. $\lambda\ll L$)\footnote{If the medium were not time-stationary, i.~e. if the wave equation depended on time, the method would also require for the characteristic time of the variations in time of the medium to be large compared to the period of the wave.}~\cite{tracy_ray_2014}. The benefit of this approximation is that the wave equation Eq.~\eqref{Eq:wave_eq}, which again is a PDE system and is hence in general difficult to solve, can be reduced as we will show to a system of ordinary differential equations (ODEs). This system, which is easier to solve, is called the ray tracing equations (RTEs)~\cite{tracy_ray_2014}. On the other hand, a consequence of this simplification is that wave optics effects such as diffraction are no longer accounted for.

The solutions of the RTEs are parametric curves in phase space, which can be interpreted as light rays in physical space. Precisely, in physical space, rays are the integral curves of the Poynting vector field~\cite{sluijter_general_2008}, or equivalently the group velocity field. They are therefore the trajectories of the wave packets~\cite{venaille_ray_2023}. Computing a large number of rays allows one to reconstruct a wave field as an approximate solution to the wave equation~\cite{tracy_ray_2014}.

As indicated above, this method boils down to the assumption of a scale separation between the dynamics of the phase and of the envelope of the wave. The former is supposed to evolve faster than the latter, whose variations are of the same order as those of the properties of the medium. A parameter $\epsilon=\lambda/L$~\cite{perez_manifestation_2021} is introduced to keep track of this ordering, with $\epsilon\ll 1$. This parameter is analogous to the Planck constant $\hbar$ of quantum physics~\cite{littlejohn_geometric_1991,venaille_ray_2023} and is therefore sometimes called the semi-classical parameter~\cite{venaille_ray_2023}. It is also called the eikonal parameter~\cite{tracy_ray_2014} or the ordering parameter~\cite{littlejohn_geometric_1991}.

The existence of this scale separation is reflected in models through the introduction of a particular ansatz for the wave field, specifically in the form of a quasi-plane wave. The envelope of such a wave  varies slowly compared to the dynamical phase, so that the wave is locally plane~\cite{tracy_ray_2014}. Specifically, we write each propagating mode supported by the medium as~\cite{tracy_ray_2014,perez_manifestation_2021}
\begin{align}\label{eq:WKB}
    \vec\Psi_m(\vec x,t)=A(\vec x) \exp\left[{i(\epsilon^{-1}\Phi_{0}(\vec x,t)+\Phi_{1}(\vec x))}\right]\uvec e(\vec x)
\end{align}
where the subscript $m$ refers to the mode $m$, and $A$, $\Phi_0$ and $\Phi_1$ are three real scalar fields which represent respectively the real amplitude, the dynamical (or eikonal) phase and the amplitude phase. $\uvec e$ is a complex unit vector field representing the polarization of the mode. The fields $A$, $\Phi_1$ and $\uvec e$ constitute the envelope of the wave and are therefore assumed to have slow variations compared to the dynamical phase $\Phi_0$. This is modelled by the factor $\epsilon^{-1}$ multiplying $\Phi_0$. Eq.~\eqref{eq:WKB} is known as the eikonal approximation~\cite{tracy_ray_2014} or the WKB ansatz~\cite{littlejohn_geometric_1991,perez_manifestation_2021}. Note that we intentionally dropped for simplicity the subscript $m$ on variables on the right hand side of Eq.~\eqref{eq:WKB}, but they are in principle all mode-dependent.

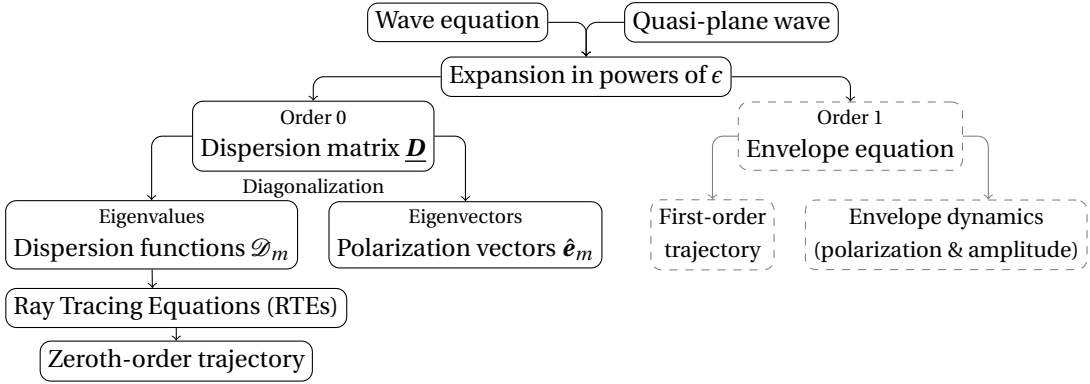
\begin{figure}
    \begin{tikzpicture}[node distance=0.5cm,anchor=west,rounded corners,rectangle,text centered]
        \node[align=center,draw=black] (wave_eq) [xshift=-0.5cm,yshift=0cm,anchor=east]
            {Wave equation};
        \node[align=center,draw=black] (quasi-plane_wave) [xshift=0.5cm,yshift=0cm,anchor=west]
            {Quasi-plane wave};
        \node[align=center,draw=black] (expansion)         [xshift=0cm,yshift=-0.45cm,anchor=north] 
            {Expansion in powers of $\epsilon$};
        \node[align=center,draw=black] (order0)            [xshift=-2cm, yshift=-1.5cm,anchor=east] 
            {{\footnotesize Order 0}\\Dispersion matrix $\tensor D$};
        \node[align=center,draw=gray,dashed] (order1)            [xshift=2cm, yshift=-1.5cm,anchor=west]
            {{\footnotesize Order 1}\\Envelope equation};
        \node[align=center] (diag)            [below of=order0,yshift=-0.18cm]   
            {\footnotesize Diagonalization};
        \node[align=center,draw=black] (disp_func)         [below of=order0,xshift=-0.2cm, yshift=-0.8cm,anchor=east]
            {{\footnotesize Eigenvalues}\\Dispersion functions $\cald_m$};
        \node[align=center,draw=black] (pola)              [below of=order0,xshift=0.2cm, yshift=-0.8cm,anchor=west]
            {{\footnotesize Eigenvectors}\\Polarization vectors $\uvec e_m$};
        \node[align=center,draw=black] (RTEs)              [below of=disp_func,anchor=north,xshift=0.325cm,yshift=-0.2cm] 
            {Ray Tracing Equations (RTEs)};
        \node[align=center, draw=black] (traj) [below of=RTEs, yshift=-0.2cm] {Zeroth-order trajectory};
        \node[align=center,draw=gray,dashed] (env)         [below of=order1,xshift=-0.6cm, yshift=-0.8cm,anchor=west]
            {\small Envelope dynamics\\\small (polarization \& amplitude)};
        \node[align=center,draw=gray,dashed] (trajo1)         [below of=order1,xshift=-1cm, yshift=-0.8cm,anchor=east]
            {\small First-order\\\small trajectory};
        \draw[->] (wave_eq) -| (expansion);
        \draw[->] (quasi-plane_wave) -| (expansion);
        \draw[->] (expansion) -| (order0);
        \draw[->] (expansion) -| (order1);
        \draw[->] (order0) -| (disp_func);
        \draw[->] (order0) -| (pola);
        \draw[->] (disp_func) -- (node cs:name=RTEs,angle=140.9);
        \draw[->] (RTEs) -- (traj);
        \draw[->,gray] (order1) -| (trajo1);
        \draw[->,gray] (order1) -| (node cs:name=env,angle=40);
    \end{tikzpicture}
    \caption{Diagram depicting the steps in the GO framework going from the wave equation to the RTEs to zeroth order. In this study we limit ourselves to the zeroth-order expansion.}
    \label{fig:GO_diagram}
\end{figure}

Plugging the ansatz Eq.~\eqref{eq:WKB} in the wave equation Eq.~\eqref{Eq:wave_eq} gives a series in powers of $\epsilon$, from which wave equations at successive orders in $\epsilon$ can be identified. This is illustrated in Fig.~\ref{fig:GO_diagram}. The leading order in $\epsilon$, which corresponds to the $\calo(\epsilon^0)$ terms, gives the zeroth-order RTEs from which are obtained the trajectory of the rays in phase space, i.~e. the position and the wavevector $(\vec x(s),\vec k(s))$ as functions of a ray parameter $s$~\cite{tracy_ray_2014}. It is analogous to the classical limit $\hbar\to 0$ in quantum physics. To obtain information on the evolution of the envelope, i.~e. of the amplitude and the polarization, along the ray, one needs to consider the RTEs to first order $\calo(\epsilon)$~\cite{tracy_ray_2014}. First-order RTEs also give first-order corrections to the trajectory describing the bending of the rays due to polarization state through spin-orbit interactions~\cite{bliokh_spinorbit_2015}, including spin-Hall effect~\cite{bliokh_geometrodynamics_2009,Fu2023} or optical Magnus effect~\cite{bliokh_modified_2004}. In this study we are however interested in the trajectory to zeroth order, and thus focus only on the zeroth-order RTEs.

\subsection{Ray tracing equations for the ray trajectory}

The wave equation to zeroth order in $\epsilon$ can be written as 
\begin{equation}\label{eq:disp_matrix}
\tensor D(\vec x,\vec k,\omega)\vec\Psi=\vec 0
\end{equation}
where we have defined the zeroth-order wavevector $\vec k$ and the angular frequency $\omega$ as
\begin{subequations}
\begin{align}
\vec k&\doteq -\vec\nabla\Phi_0\\
\omega&\doteq\partial_t\Phi_0,
\end{align}
\end{subequations}
and where $\tensor D$ is a matrix called the dispersion matrix~\cite{littlejohn_geometric_1991}. This matrix describes the local behaviour of the wave. Indeed, since it is to zeroth order in $\epsilon$, it does not include any derivative of the medium properties. As shown in Fig.~\ref{fig:GO_diagram}, the eigenvalues of this matrix, which we write $\cald_m$, are the zeroth-order dispersion functions, and the corresponding unit eigenvectors $\uvec e_m$ are the unit polarization vectors of the modes~\cite{tracy_ray_2014}. The determinant of $\tensor D$ then writes 
\begin{equation}
\cald \doteq \det(\tensor D) = \prod_m\cald_m.
\end{equation}

Eq.~\eqref{eq:disp_matrix} has non-trivial solutions if and only if the full dispersion relation $\cald(\vec x,\vec k,\omega)=0$ holds. This requires that $\cald_m(\vec x,\vec k,\omega)=0$ is verified for at least one mode $m$, which is then known as the dispersion relation for the mode $m$. If it is satisfied the mode $m$ propagates with a non-zero amplitude. The dispersion relation for a given mode $m$ can be viewed as the Hamilton-Jacobi equation of a particle with position $\vec x$ and momentum $\vec k$, and $\cald_m$ the Hamiltonian~\cite{littlejohn_geometric_1991}. Using this analogy, one writes the equations for the trajectory in phase space as Hamilton's equations~\cite{tracy_ray_2014}
\begin{subequations}\label{eq:traj_RTEs}
    \begin{align}
        \fracd{\vec x}{s}&=-\vec\nabla_{\vec k}\cald_m\\
        \fracd{\vec k}{s}&=\vec\nabla_{\vec x}\cald_m.
    \end{align}
\end{subequations}
We call these equations the trajectory RTEs. Their solutions are the trajectories of the rays in phase space $(\vec x(s),\vec k(s))$ for the mode $m$~\cite{tracy_ray_2014}. These trajectories are those of the energy of the wave, i.~e. of the wave packet or, with a quantum physics point of view, those of the photons. These equations are similar to those of the dynamics of a point-particle~\cite{ruiz_first-principles_2015}. The ray in physical space is tangent to the group velocity $\vec v_g\doteq d{\vec x}/{dt}$. This can be shown from Eq.~\eqref{eq:traj_RTEs} through a reparametrization of the ray using the time $t$ instead of $s$. Again, since these equations are to zeroth order in $\epsilon$, they do not provide any information on the evolution of the amplitude or of the polarization. 

Note importantly here that within this formalism computing the trajectory of a given mode $m$ only requires to have the dispersion function $\cald_m$ describing this mode. In the next section we will therefore aim at obtaining these functions, playing the role of Hamiltonians, for modes propagating in moving anisotropic media.


\section{Dispersion function for a moving media}
\label{Sec:SecIII}

In this section we proceed with the derivation of wave dispersion function for a moving medium, starting with the simpler case of a medium in uniform linear motion, and then generalizing to arbitrary motions.

\subsection{Electromagnetism in uniformly moving media}\label{sub:uniform_velocity}

Let us start by showing how to relate the properties of wave propagation in a medium in uniform linear motion to a laboratory frame dispersion function describing the local behaviour of the wave.

\subsubsection{Lorentz transformations}

We begin by classically defining two reference frames : the frame $\Sigma$ attached to the laboratory in which the observer is at rest, called the lab-frame, and the frame $\Sigma'$ attached to the uniformly moving medium, called the rest-frame, in which the medium is at rest~\cite{lee_electromagnetic_1964,lee_radiation_1966}. All quantities expressed in the rest-frame $\Sigma'$ will be indicated by a prime. Let us also write $\vec v$ the constant velocity of the medium.

If we assume the lab-frame $\Sigma$ to be inertial then $\Sigma'$ is also inertial and the Lorentz transformations\footnote{To be exact, here we use the Lorentz transformations for a Lorentz boost.} can be used to transform the position four-vector $x^\mu=(ct,\vec x)$\footnote{Using a common notation abuse, we write here the contravariant (resp. covariant) four-vectors by the notation of their contravariant (resp. covariant) components where the Greek indices go from 0 to 3.} and the four-wavevector $k^\mu=(\omega/c,\vec k)$ from one frame to the other. The Lorentz transformations for these four-vectors respectively write~\cite{mccall_relativity_2007}
\begin{subequations}\label{eq:Lorentz_transfo_xt}
    \begin{align}\label{eq:relat_Doppler_xt}
        ct'&=\gamma(ct-\vec\beta\cdot\vec x)\\
        \vec x'&=(\eye+(\gamma-1)\vec\beta\otimes\vec\beta/\beta^2)\vec x-\gamma\vec\beta ct,
    \end{align}
\end{subequations}
and 
\begin{subequations}\label{eq:Lorentz_transfo_omegak}
    \begin{align}\label{eq:relat_Doppler_omega}
        \omega'&=\gamma(\omega-c\vec\beta\cdot\vec k)\\\label{eq:relat_aberration}
        \vec k'&=(\eye+(\gamma-1)\vec\beta\otimes\vec\beta/\beta^2)\vec k-\gamma\vec\beta \omega/c
    \end{align}
\end{subequations}
where the vector $\vec\beta=\vec v/c$ is the dimensionless velocity, $\gamma={1}/{\sqrt{1-\beta^2}}$ is the Lorentz factor and $\otimes$ is the tensor product. The first equation of the Lorentz transformation for the four-wavevector, Eq.~\eqref{eq:relat_Doppler_omega}, captures the relativistic Doppler effect~\cite{mccall_relativity_2007,censor_dispersion_1980} which causes a shift in the frequency of the wave going from one frame to the other. The second equation of the Lorentz transformation for the four-wavevector, Eq.~\eqref{eq:relat_aberration}, describes the relativistic aberration effect, that is the change in the direction of the wavevector going from one frame to the other. Both the Doppler effect and the aberration effect depend on the direction of the wavevector $\vec k$, and notably the angle it makes with the velocity $\vec v$, and on the wave frequency.

\subsubsection{Effective medium}

\begin{figure}
    \begin{tikzpicture}[node distance=0.8cm]
        \node (medium_at_rest) [xshift=0cm] {\includegraphics[width=1cm]{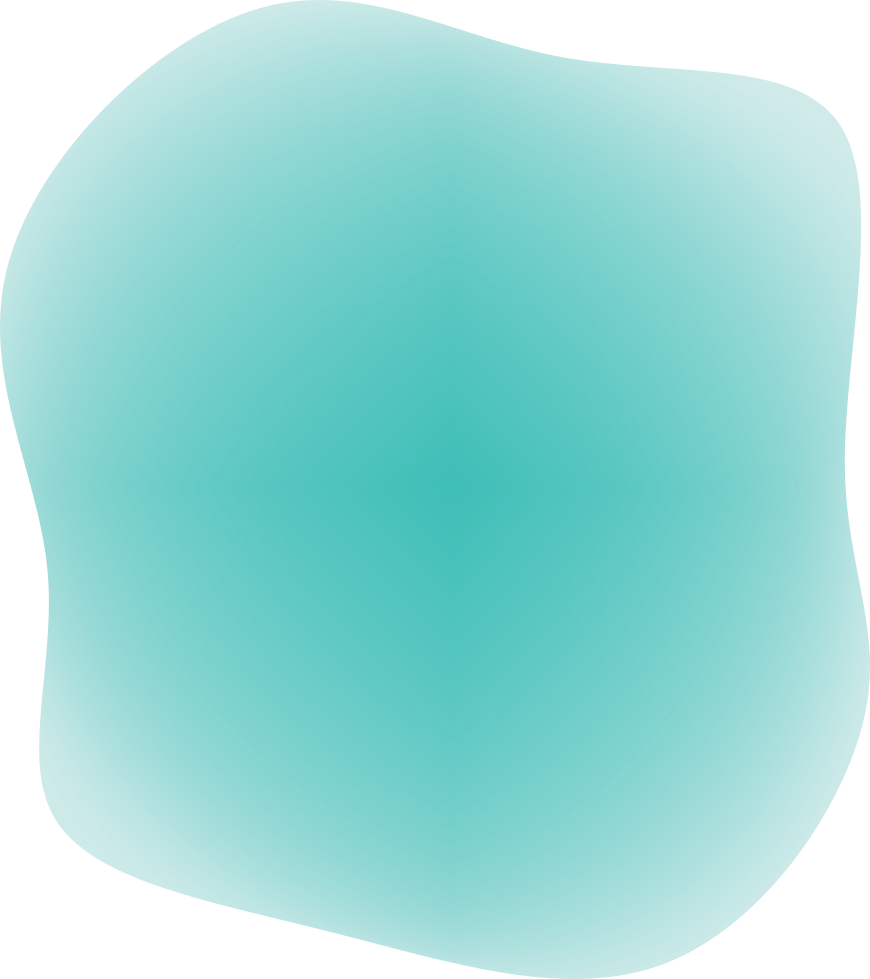}};
        \node[align=right] (medium_at_rest_text) [left of=medium_at_rest,anchor=east] {\bfseries Medium\\\bfseries at rest};
        \node (cald') [xshift=0cm] {$\cald'_m$};
        \node (+)  [below of=medium_at_rest,xshift=0cm] {\Huge $+$};
        \node (motion) [below of=+] {\includegraphics[width=1cm]{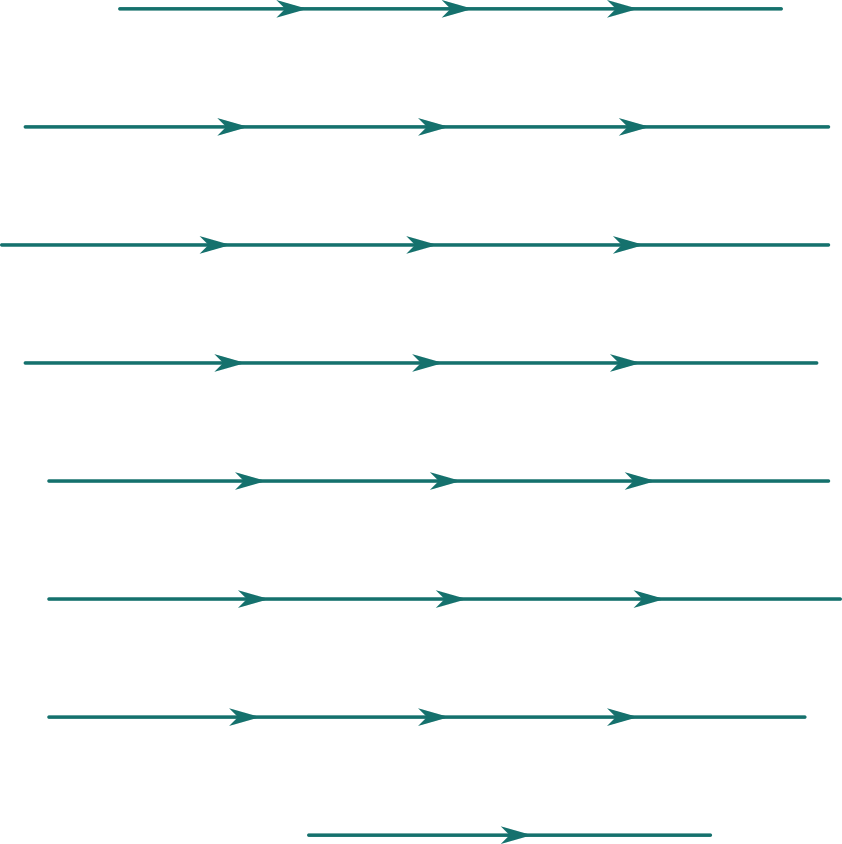}};
        \node (motion_text) [left of=motion,anchor=east]{\bfseries Motion};
        \node (equal)  [right of=+,xshift=0.4cm] {\Huge $=$};
        \node (moving_medium) [right of= equal,xshift=1cm] {\includegraphics[width=2cm]{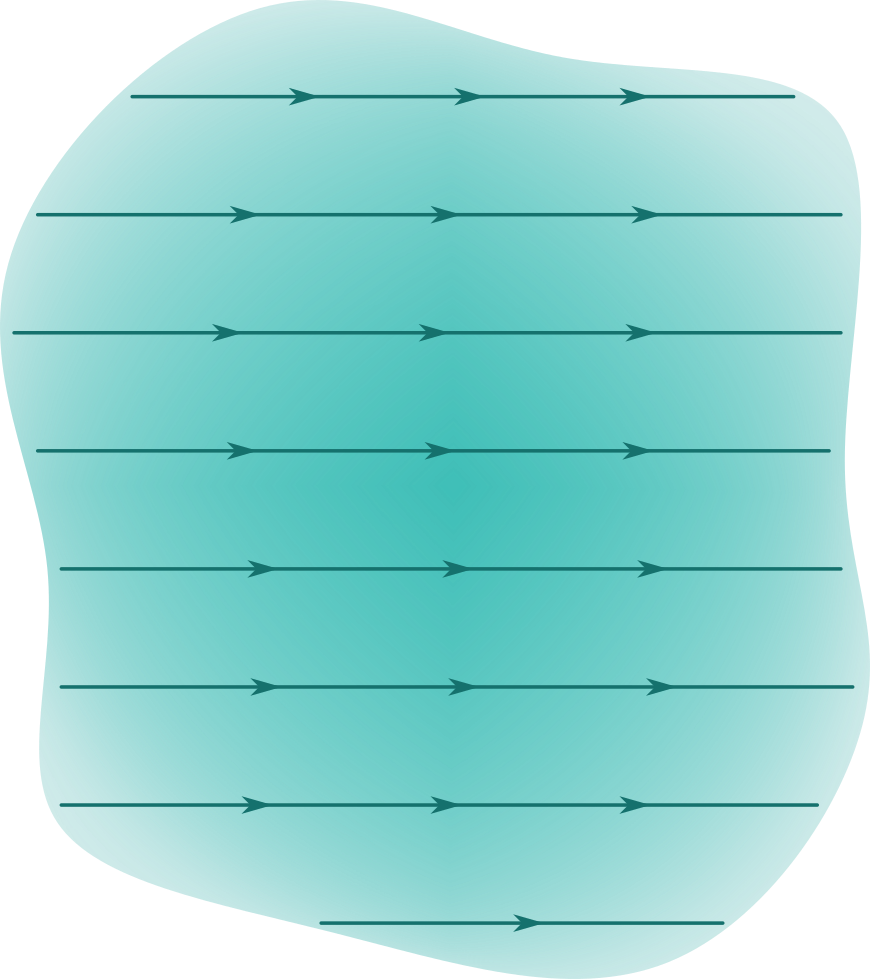}};
        \node (moving_medium_text) [below of= moving_medium, yshift=-0.5cm,anchor=north] {\bfseries Moving medium};
        \node (equivalent) [right of=moving_medium,xshift=1cm] {\Huge $\Leftrightarrow$};
        \node (effective_medium) [right of= equivalent,xshift=1cm] {\includegraphics[width=2cm]{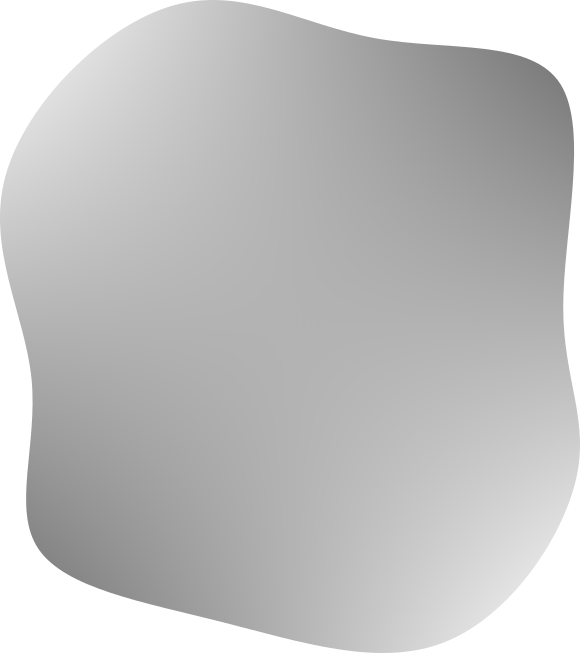}};
        \node[align=center] (effective_medium_text) [below of= effective_medium, yshift=-0.5cm,anchor=north] {\textbf{Equivalent effective}\\\textbf{medium at rest}};
        \node (cald) [right of= equivalent,xshift=1cm] {$\cald_m$};
        \draw[->] (node cs:name=medium_at_rest,angle=35) ..controls +(2,0.5) and +(-1,0.3) .. node[above,align=right] {\textit{Emergence of additional properties}\\\textit{due to motion}} (node cs:name=effective_medium,angle=140);
    \end{tikzpicture}
    \caption{The moving medium is modelled in the lab-frame as an equivalent effective medium with additional properties that result from motion. Propagation in this effective medium is described by the dispersion function $\cald_m$, which differs from the dispersion function $\cald'_m$ known in the original medium at rest.}
    \label{fig:Fig2}
\end{figure}
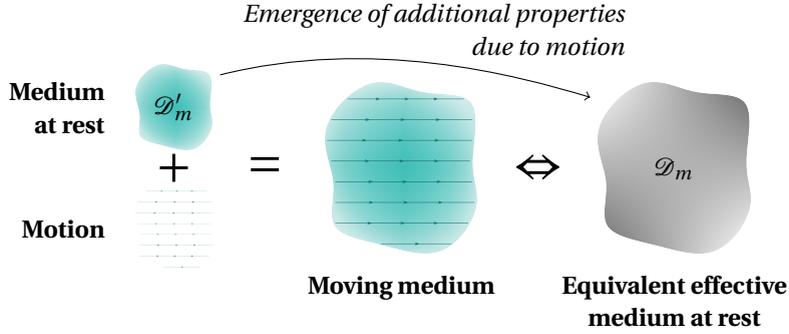

Our goal is to describe the propagation of waves as observed by an observer at rest in the lab-frame. The approach adopted here, as depicted in Fig.~\ref{fig:Fig2}, is to consider the moving medium as an equivalent effective medium at rest in the lab-frame, but with additional properties due to its motion~\cite{lopez_dispersion_1996,lopez_dispersion_2004}. The velocity is then seen as a vector property of the effective medium, in a way like the magnetic field.

If we denote $\cald'_m$ the dispersion functions of the modes propagating in the medium when it is at rest, i.~e. in the frame $\Sigma'$, then let $\cald_m$ be the dispersion functions of these same propagating modes but seen from the lab-frame $\Sigma$ where the medium is moving. The $\cald_m$ are therefore the dispersion functions describing the propagation of the modes in the effective medium~\cite{censor_relativistic_2010}. From there, the effective medium can be treated as any classical medium at rest. The effective medium is, however, more complicated than the original medium since the velocity adds a preferred direction. The effective medium is then in general bianisotropic~\cite{cheng_covariant_1968,mccall_relativity_2007}, even if the original medium is isotropic, and spatially dispersive~\cite{lopez_dispersion_1996} even if the original medium is only time dispersive. Yet, it is worth noting that the effective anisotropy due to motion does not necessarily lead to an effective birefringence~\cite{mccall_relativity_2007}. For example, in the case of a moving isotropic medium, the two modes seen from the lab-frame are degenerate, i.~e. they have the same dispersion relation, although the effective medium is bianisotropic due to motion.

\subsubsection{Dispersion function in the lab frame}

There are at least two ways to obtain the dispersion functions $\cald_m$ of the effective medium that are needed to use the RTEs Eq.~\eqref{eq:traj_RTEs}, and from there to compute the trajectories of the rays~\cite{mccall_relativity_2007,lopez_dispersion_1996}.

The first one, used notably in Refs.~\cite{tai_study_1964,tai_electrodynamics_1965,lee_radiation_1966,gotte_dragging_2007,gueroult_determining_2019}, is what we call the \textit{lab-frame approach} since the main part of the calculations are done with a lab-frame point of view. The first step is to derive the constitutive relations of the effective medium as seen in the lab-frame. For this one typically starts with the constitutive relations in the rest-frame before invoking the Lorentz transformations of the electromagnetic fields~\cite{minkowski_grundgleichungen_1910} to rewrite these relations in terms of the lab-frame electromagnetic fields. These relations are finally combined with Maxwell's equations written in the lab-frame to obtain a wave equation in the lab-frame. Proceeding in the framework of GO as explained in Section~\ref{Sec:SecII}, the lab-frame dispersion matrix $\tensor D$ is then obtained from the wave equation to zeroth order in the small eikonal parameter $\epsilon$, and the dispersion functions $\cald_m$ of the modes are the eigenvalues of this matrix. While conceptually straightforward, this approach has two major drawbacks. First, finding analytical expressions for the eigenvalues and eigenvectors of the dispersion matrix is only tractable for simple media. Second, the calculations become very complicated when one wants to work beyond the first order in $\beta$. To circumvent these difficulties, we will generally prefer to use the second approach introduced below.

The second approach, adopted in~\cite{unz_relativistic_1968,unz_relativistic_1966,meyer-vernet_high-frequency_1980} and referred to as the \textit{transformation approach}, uses the fact that the dispersion relation is Lorentz covariant, in the sense that the dispersion relation in one frame can be obtained from the one in the other frame only by Lorentz transforming the position four-vector $x^\mu=(ct,\vec x)$ and the four-wavevector $k^\mu=(\omega/c,\vec k)$ using the Lorentz transformations Eqs.~\eqref{eq:Lorentz_transfo_xt} and \eqref{eq:Lorentz_transfo_omegak}. This result was demonstrated by Censor both for homogeneous anisotropic media~\cite{censor_dispersion_1980} and bianisotropic media~\cite{censor_relativistic_2010}. Following Refs.~\cite{censor_dispersion_1980,censor_relativistic_2010}, we write this property as
\begin{equation}\label{eq:D=D'}
    \cald_m(x^\mu,k^\mu)=\cald'_m(x'^{\mu}(x^\mu),k'^\mu(k^\mu))
\end{equation}
where $x'^\mu$ and $k'^\mu$ are written as functions of $x^\mu$ and $k^\mu$ through the Lorentz transformations Eqs.~\eqref{eq:Lorentz_transfo_xt} and \eqref{eq:Lorentz_transfo_omegak}. This amounts in fact to taking into account the relativistic Doppler shift and the aberration effect seen in the rest-frame due to the medium's velocity. Thanks to Eq.~\eqref{eq:D=D'}, one only needs to know the dispersion function in the rest-frame $\cald'_m$, which is often the case, to immediately obtain the dispersion function of the effective medium $\cald_m$ in the lab-frame. An advantage of this approach compared to the first one is that it works for very general media~\cite{censor_relativistic_2010}. This is notably true for dispersive anisotropic media like magnetized plasmas~\cite{unz_relativistic_1968}, as we will discuss in Section~\ref{Sec:SecV}. Another advantage is that it yields the dispersion functions to all orders in $\beta$, hence allowing for relativistic velocities. On the other hand, a limit of this second approach is that it cannot give the polarization vectors nor the envelope equation since it bypasses the derivation of a wave equation, and from there of a lab-frame dispersion matrix. It is only effective to find the dispersion function from the lab-frame.

An important caveat which we must stress here is that we assume the covariance property to hold for inhomogeneous media (as shown by the $x^\mu$ dependency in Eq.~\eqref{eq:D=D'}), whereas it was only demonstrated by Censor for homogeneous media~\cite{censor_relativistic_2010}. Note also that Censor's covariant argument referred to the full dispersion relation $\cald(x^\mu,k^\mu)=0$, whereas we use it here for each mode $m$. This splitting will however be shown to hold at least for the modes considered in Section~\ref{Sec:SecV}. 

\subsubsection{Illustration for homogeneous isotropic media}

Let us now briefly illustrate this transformation approach by going back to our example of a homogeneous isotropic non-dispersive medium, which we now consider to be moving. The refractive index of this medium at rest is therefore denoted $n'$. The rest-frame dispersion function of the modes propagating in such a medium is well-known, namely
\begin{equation}
\cald'_\text{iso}=n'^2\frac{\omega'^2}{c^2}-k'^2.
\end{equation}
Using the Lorentz transformations Eqs.~\eqref{eq:Lorentz_transfo_omegak}, we then get from covariance the dispersion function in the lab-frame~\cite{leonhardt_optics_1999}
\begin{equation}\label{eq:iso_disp_func}
    \cald_\text{iso}=\frac{\omega^2}{c^2}-k^2+(n'^2-1)\gamma^2 \left(\frac{\omega}{c}-\vec\beta\cdot\vec k\right)^2.
\end{equation}
We see that $\cald_\text{iso}$ now depends on the direction of $\vec k$ through the term $\vec\beta\cdot\vec k$, whereas it was not the case for $\cald'_\text{iso}$. Note also that for the vacuum ($n'=1$) the dispersion functions in the two frames are the same, which is consistent with the fact that a moving vacuum remains a vacuum. This comes from the fact that the quantity $\omega'^2/c^2-k'^2=\omega^2/c^2-k^2$ is a Lorentz scalar, and is thus invariant under Lorentz transformations.\\

Summing our findings, we have identified a way to obtain the dispersion function $\cald_m$ necessary to use the trajectory RTEs in the lab-frame, under the requirement that the rest-frame dispersion function $\cald'_m$ is known. It is worth noting that to study all the modes propagating in a moving medium, one has to consider also the purely oscillatory modes and the evanescent modes which can become propagative from lab-frame although they do not propagate in the rest-frame~\cite{censor_dispersion_1980}. This behaviour, which we interpret as a consequence of the loss of simultaneity between the two reference frames according to relativity, will be illustrated in Section~\ref{Sec:SecV}.

\subsection{Dispersion function for arbitrarily moving media}\label{sub:arbitrary_velocity}

Having shown in the previous paragraph how to obtain lab-frame dispersion function for a medium in uniform linear motion, we would like to extend it to the case of a medium in arbitrary motion, as considered in Refs.~\cite{leonhardt_optics_1999,bourgoin_relativistic_2021}.  We argue here that this generalization in fact comes naturally in the geometrical optics approximation, assuming weak enough spatial inhomogeneity in the velocity field. 

To see this, let us consider a medium with an arbitrary time-stationary\footnote{The stationary hypothesis is only here to simplify the discussion by having a wave equation independent in time, as indicated in the general introduction of GO at the beginning of Section~\ref{Sec:SecII}.} velocity field $\vec \beta(\vec x)=\vec v(\vec x)/c$. The idea explored here is to take advantage of the scale separation that is as shown in Section~\ref{Sec:SecII} characteristic of the GO framework. Following this direction, we now require for the characteristic variation length of the velocity field to be large compared to the wavelength~\cite{leonhardt_optics_1999,censor_ray_1976}. As a result $\vec\beta$ does not change significantly over a wavelength, and the wave thus sees locally a medium in uniform linear motion. In this limit, and assuming that "special relativity theory is valid locally and instantaneously"~\cite{censor_ray_1976}, the local behaviour of the wave is then found to be described by $\cald_m(\vec x,\vec k,\omega)$~\cite{venaille_ray_2023} as obtained using the methods for a uniform linear motion introduced in Subsection~\ref{sub:uniform_velocity}, considering crucially the local velocity $\vec\beta(\vec x)$ as a constant~\cite{leonhardt_optics_1999}. Following this route, we get $\cald_m(\vec x,\vec k,\omega)$ at each position $\vec x$. This lab-frame dispersion function $\cald_m(\vec x,\vec k,\omega)$ does not contain derivatives of the velocity field, as expected to zeroth order in the small eikonal parameter $\epsilon$.

Note that while, thanks to the geometrical optics framework used here, a non-uniform motion can be modelled via the effects known for a uniform motion, a non-uniform motion adds new difficulty in that determining the rest-frame dispersion function may be more intricate. This is because the constitutive relations in the rest-frame of an accelerated media differ from those in this same media at rest. This behaviour has long been known for rotations~\cite{Heer1964,Anderson1969,shiozawa_phenomenological_1973}, and has recently been clarified in rotating plasmas~\cite{langlois_contribution_2023}.

\section{Ray tracing in arbitrarily moving media}
\label{Sec:SecIV}
\label{sub:RTE_arbitrary_velocity}

Having identified in Section~\ref{Sec:SecIII} a framework to derive the lab-frame dispersion functions $\cald_m$ describing the modes in a medium in arbitrary motion, we can now use them in the trajectory RTEs Eqs.~\eqref{eq:traj_RTEs} derived in Section~\ref{Sec:SecII} to compute the trajectory of the rays, and identify the effect of motion. 

\subsection{RTEs for a homogeneous isotropic media in arbitrary motion}

Here we illustrate the effect of motion on rays by considering again the example of a moving homogeneous isotropic non-dispersive medium. Using the lab-frame dispersion function Eq.~\eqref{eq:iso_disp_func} and limiting ourselves here to first order in $\beta$ for the sake of simplicity, we find the trajectory RTEs\begin{subequations}\label{eq:RTEs_iso}
    \begin{align}\label{eq:dxds_iso}
        \fracd{\vec x}s&=2\vec k+2(n'^2-1)\frac{\omega}{c}\vec\beta ,\\
        \fracd{\vec k}s&=2(n'^2-1)\frac \omega c\left[(\vec\nabla\times\vec\beta)\times\vec k-(\vec k\cdot\vec\nabla)\vec\beta\right].\label{eq:dkds_iso}
    \end{align}
\end{subequations}

\begin{figure}
  \hspace{0.03\linewidth}
  \begin{subfigure}{0.44\linewidth}
    \centering
    \includegraphics[width=0.5\linewidth]{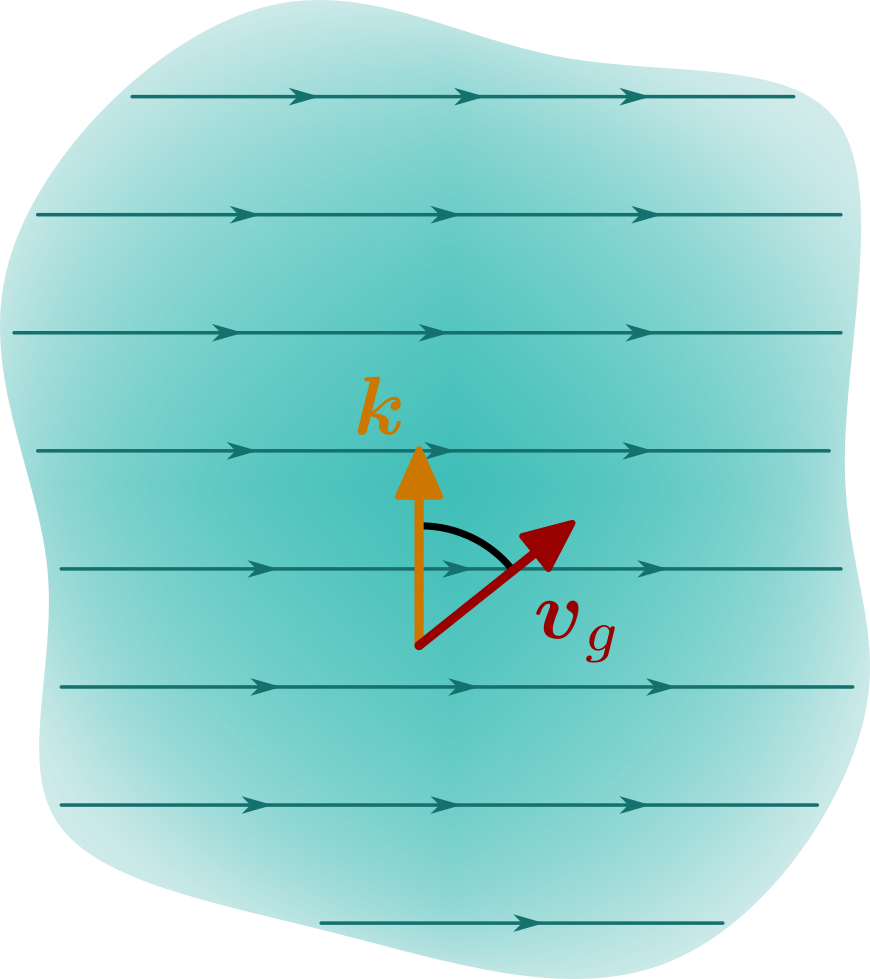}
    \caption{Illustration of the angle between the wavevector $\vec k$ and the group velocity $\vec v_g$ revealing the drag undergone by the wave due to the motion of the medium.}
    \label{subfig:drag_angle}
  \end{subfigure}
  \hspace{0.03\linewidth}
  \begin{subfigure}{0.44\linewidth}
    \centering
    \includegraphics[width=0.5\linewidth]{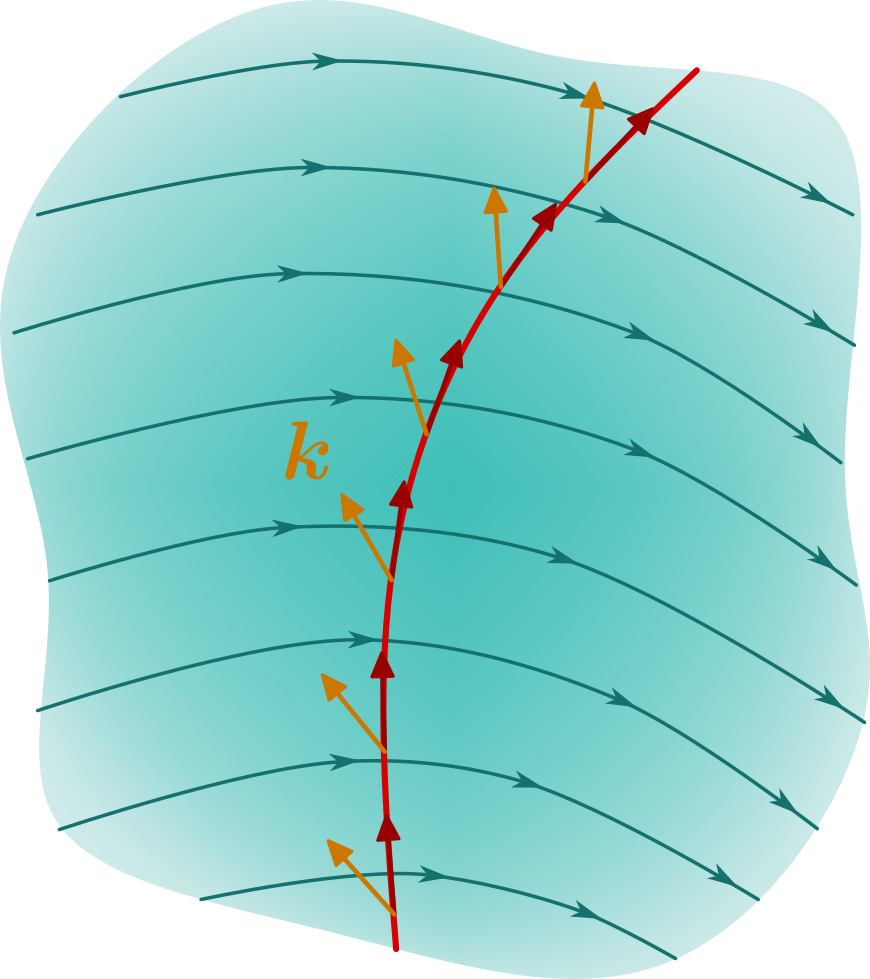}
    \caption{Illustration of the bending of the ray due to the variation of the wavevector $\vec k$ caused by the non-uniformity of the velocity field $\vec v(\vec x)$ of the medium.}
    \label{subfig:bended_ray}
  \end{subfigure}
  \hspace{0.03\linewidth}
  \caption{Illustrations of some effects of the motion on the trajectory of the wave.}
\end{figure}

Examining Eqs.~\eqref{eq:RTEs_iso}, we first note that the velocity appears in Eq.~\eqref{eq:dxds_iso} in the form of a term proportional to $\vec\beta$ on the right hand side, which supplements the term $2\vec k$ classically expected without motion. This term modifies the group velocity $\vec v_g\propto d{\vec x}/{ds}$ of the wave as seen in the lab-frame. Specifically, there is now as illustrated in Fig.~\ref{subfig:drag_angle} a finite angle between the wavevector $\vec k$ and the group velocity $\vec v_g$, in contrast with what is found in this isotropic medium at rest where those two vectors are aligned. Considering the particular case $\vec k\perp\vec\beta$ for which from Eq.~\eqref{eq:iso_disp_func} $n'=n$, one finds from Eq.~\eqref{eq:dxds_iso} that the lateral shift $\delta$ per unit length along $\vec k$ induced by the motion writes
\begin{equation}
\delta = \left(n'-\frac{1}{n'}\right)\frac{v}{c},
\end{equation}
which is precisely the lateral shift predicted by Player~\cite{player_dispersion_1975} and observed by Jones~\cite{jones_fresnel_1972,jones_aether_1975} when neglecting dispersion.

Besides the light dragging seen in Eq.~\eqref{eq:dxds_iso}, we see that the motion also appears in the second equation Eq.~\eqref{eq:dkds_iso}, though it now appears through the non-uniformity of the velocity field. This was to be expected in that the velocity field non-uniformity leads to inhomogeneous properties for the effective medium. Similarly to a refractive index non-uniformity, the velocity field non-uniformity now causes a bending of the rays, affecting the evolution of $\vec k$ along the ray. This second manifestation of motion is illustrated in Fig.~\ref{subfig:bended_ray}.

\subsection{Relation to previous work}

Let us now discuss how the model for ray tracing in moving media proposed above, that is the combination of the trajectory RTEs Eqs.~\eqref{eq:traj_RTEs} and the lab-frame dispersion function $\cald_m$ obtained by Lorentz transforming the rest-frame one $\cald'_m$, relates to previous contributions on the topic.

Examining the particular case of an isotropic non-dispersive medium, Rozanov~\cite{rozanov_geometrical_2005} derived from the wave equations for the electric and magnetic fields, combined with the expression of the Poynting vector, a different set of RTEs, valid to first order in $\beta$. 
In this model the effects of motion then appears as a perturbation term in the trajectory RTEs, which is a correction to the result for zero motion. This perturbation term can be interpreted, under the point-particle analogy, as an additional force term which deviates the trajectory of the rays. We verified that the trajectory RTEs Eqs.~\eqref{eq:traj_RTEs} equipped with the lab-frame dispersion function $\cald_m$, are equivalent to the trajectory RTEs with the perturbation term found by Rozanov~\cite{rozanov_geometrical_2005}.

Still for isotropic non-dispersive medium, another approach that has been explored is to make use of a relativistic formalism~\cite{leonhardt_optics_1999,gordon_zur_1923,bourgoin_general_2020,bourgoin_relativistic_2021}, valid this time to all orders in $\beta$. Specifically, these contributions make use of the fact that the covariant constitutive relations for a moving isotropic medium can be recast as famously done by Gordon~\cite{gordon_zur_1923} into the metric of an empty curved spacetime. This equivalent virtual spacetime is called optical spacetime and is equipped with an optical metric $\bar g$, which depends on the characteristics of the medium and on its motion.
Using this analogy, wave propagation properties in the moving medium in physical spacetime are then the same as those derived in an adequately curved empty spacetime. 
Specifically, the dispersion function writes~\cite{gordon_zur_1923,bourgoin_general_2020,bourgoin_relativistic_2021}
\begin{align}
    \cald_\text{Gordon}&=\bar g^\munu k_\mu k_\nu&\text{with}&&\bar g^\munu&=g^\munu+(n'^2-1)\beta^\mu\beta^\nu,
\end{align}
where $\bar g^{\mu\nu}$ is the Gordon optical metric, $g^{\mu\nu}$ is the metric of the physical space and $\beta^\mu=\gamma(1,\vec\beta)$ is the dimensionless four-velocity of the medium. One verifies that this dispersion function is precisely the one identified above in Eq.~\eqref{eq:iso_disp_func} when using Minkowski's metric $\eta^{\mu\nu}=\operatorname{diag}(+,-,-,-)$ as the background metric $g^{\mu\nu}$.

In short, the theory proposed here is found to be consistent with known results previously obtained in the limit of isotropic dielectrics. Compared to these earlier contributions though, a strong advantage of the proposed approach, and principal motivation for its development, is that it can accommodate dispersive and anisotropic media, such as plasmas as considered in the next Section.

\section{Ray tracing in moving plasmas}\label{Sec:SecV}

As an illustration of the method, the theory established in Sections ~\ref{Sec:SecIII} and \ref{Sec:SecIV} is now applied in this Section to a cold moving magnetized plasma. The emphasis is put on deriving the lab-frame dispersion functions $\cald_m$, as needed to compute the RTEs. Note that the problem of finding the lab-frame dispersion relations for a plasma in uniform linear motion has already received attention, both using the lab-frame approach~\cite{unz_magneto-ionic_1962,lee_radiation_1966,tai_electrodynamics_1965} and the transformation approach~\cite{unz_relativistic_1968,unz_relativistic_1966,meyer-vernet_high-frequency_1980}. These previous contributions did however focus primarily on the high-frequency electronic response. Here we 
follow the approach presented in Section~\ref{Sec:SecIII} to derive these results without any particular assumption, recovering and generalizing these earlier contributions. 

\subsection{Moving unmagnetized plasma}

As a first step we consider the simple case of an unmagnetized plasma. There are two different modes in an unmagnetized plasma at rest: the propagating electromagnetic mode and the oscillating electrostatic mode. The latter does not propagate.

\subsubsection{The propagating electromagnetic mode : the O mode}

The dispersion function of the propagating mode (ordinary or O mode) is~\cite{rax_physique_2005}
\begin{equation}
    \cald'_O(\vec k',\omega')=\omega'^2/c^2-k'^2-\omega'^2_p/c^2
\end{equation}
where $\omega'^2_p=\omega'^2_{pe}+\omega'^2_{pi}$ with $\omega_{ps}' = [n_{s}' e^2/(m_{s}'\epsilon_0)]^{1/2}$ the plasma frequency of the species $s$ as measured in the rest-frame $\Sigma'$. Recalling that the quantity $\omega'^2/c^2-k'^2$ is a Lorentz scalar, the lab-frame dispersion function writes straightforwardly
\begin{equation}
    \cald_O(\vec k,\omega)=\omega^2/c^2-k^2-\omega'^2_p/c^2.
    \label{Eq:disp_O_lab}
\end{equation}
The dispersion function of the O mode is then Lorentz invariant : it has the same form whether it is expressed in $\Sigma'$ or $\Sigma$ ~\cite{ko_passage_1978}. In particular, the dispersion function from the lab-frame $\cald_O$ has no dependence on the velocity $\vec\beta$. This is a notable feature since it means that the O mode is not affected by the motion of the plasma, as already noted by  Mukherjee~\cite{mukherjee_electromagnetic_1975} and Ko and
Chuang~\cite{ko_passage_1978}. In other words, it propagates identically whether the plasma is moving or not. This result suggests that diagnostics using O mode will likely not be affected by motion, at least in the cold plasma limit and as long as one is focusing on trajectory\footnote{There could still be first-order corrections to the RTEs, as shown for an inhomogeneous plasma in Ref.~\cite{Fu2023}, though in this case the corrections would stem from motion. }. Similarly, this results a priori prohibits using an O mode to probe the motion of a cold plasma.

\subsubsection{The oscillating electrostatic mode : Langmuir waves}

The dispersion function of the oscillating mode, also known as Langmuir waves, is~\cite{rax_physique_2005}
\begin{equation}
    \cald'_\text{osc}(\omega')=\frac{1}{c^2}\left(\omega'^2-\omega'^2_p\right).
\end{equation}
Using the Lorentz transformations Eqs.~\eqref{eq:Lorentz_transfo_omegak} to write $\omega'$ as a function of $\omega$ and $\vec k$, we then obtain the lab-frame dispersion function~\cite{censor_dispersion_1980}
\begin{equation}
    \cald_\text{osc}(\vec x,\vec k,\omega)=\gamma^2\left(\frac\omega c-\vec k\cdot\vec\beta(\vec x)\right)^2-\frac{\omega'^2_p}{c^2}.
\end{equation}
Unlike $\cald'_\text{osc}$, $\cald_\text{osc}$ contains both $\omega$ and $\vec k$. This implies that this mode is propagative. The oscillating mode, which does not propagate in the rest-frame, does propagate in the lab-frame~\cite{censor_dispersion_1980}. As mentioned above this is a consequence of the loss of simultaneity from the rest-frame $\Sigma'$ to the lab-frame $\Sigma$. Specifically, the phase of electrostatic oscillations is the same at all points in space in $\Sigma'$, that is to say that there is simultaneity. This is however no longer the case in $\Sigma$ due to special relativity, hence the emergence of a propagating wave.

\subsection{Moving magnetized plasma}

Moving to magnetized plasmas, the method presented in Section~\ref{Sec:SecIII} can be applied to obtain the dispersion functions in the lab-frame for modes propagating at an arbitrary angle with respect to the background magnetic field $\vec B_0$ in a moving magnetized cold plasma. This essentially boils down to substituting to $\omega'$ and $\vec k'$ in the classic Appleton-Hartree equation~\cite{appleton_wireless_1932}  -- known to hold in the plasma rest-frame -- their expressions in terms of the lab-frame variables $\omega$ and $\vec k$, as prescribed by Eq.~\eqref{eq:Lorentz_transfo_omegak}. Although lengthy, this task poses no difficulty.

Rather than dealing with these general formulas though, we focus here on the simpler case where the wavevector is perpendicular to the magnetic field ($\vec k'\perp\vec B'_0$). In this configuration, that is for perpendicular propagation, the modes propagating in the rest-frame are the well known O and X (for extraordinary) modes. The O mode is the same as in an unmagnetized plasma. From the discussion above we already have its dispersion function from the lab-frame, that is Eq.~\eqref{Eq:disp_O_lab}. 

Focusing on the X mode, the dispersion function in the rest-frame is~\cite{rax_physique_2005}
\begin{equation}
\label{eq:disp_x_rest}
    \cald'_X(\vec k',\omega')=\frac{\omega'^2}{c^2}-k'^2+\frac{\omega'^2}{c^2}\left[\chi'_\perp(\omega')-\frac{\chi'^2_\times(\omega')}{1+\chi'_\perp(\omega')}\right]
\end{equation}
with
\begin{subequations}
\label{Eq:Tensor_component}
\begin{align}
{\chi}'_{\perp}(\omega') & = \sum\limits_{s}\frac{\omega_{ps}'^2}{\Omega_{cs}'^2-\omega'^2}\label{Eq:Tensor_component_perp}\\ 
{\chi}'_{\times}(\omega') & = \sum\limits_{s}\epsilon_{s}\frac{\Omega_{cs}'}{\omega'}\frac{\omega_{ps}'^2}{\omega'^2-\Omega_{cs}'^2}\label{Eq:Tensor_component_cross}
\end{align}
\end{subequations}
the classical perpendicular and cross-field components of the susceptibility tensor $\tensor\chi'$ of a magnetized cold plasma. Here $\Omega_{cs}' = |q_s|B_0'/m_s'$ is the unsigned cyclotron frequency of species $s$ written in the rest-frame, and $\epsilon_{s} = q_s/|q_s|$. Using the Lorentz transformations Eqs.~\eqref{eq:Lorentz_transfo_omegak} in Eq.~\eqref{eq:disp_x_rest} directly yields the lab-frame  dispersion function
\begin{align}
\label{eq:disp_x_lab}
    \cald_X(\vec x,\vec k,\omega)=&\frac{\omega^2}{c^2}-k^2\nonumber\\
    &\quad+\gamma^2\left(\frac\omega c-\vec k\cdot\vec\beta(\vec x)\right)^2\left[\chi'_\perp\left(\gamma(\omega-c\vec k\cdot\vec\beta(\vec x))\right)-\frac{\chi'^2_\times\left(\gamma(\omega-c\vec k\cdot\vec\beta(\vec x))\right)}{1+\chi'_\perp\left(\gamma(\omega-c\vec k\cdot\vec\beta(\vec x))\right)}\right].
\end{align}
We immediately see that, in contrast with the O mode, $\cald_X$ depends on $\vec\beta$, which implies that the propagation of the X mode is modified by the plasma motion. 

We verify that Eq.~\eqref{eq:disp_x_lab} is consistent, when considering only the electronic response, with the result obtained by Unz~\cite{unz_relativistic_1968} in the particular configuration $\vec k'\perp\vec B'_0$ considered here and neglecting collisions. In the limit $\vec v\perp\vec B'_0$ it also agrees with the result derived by Mukherjee~\cite{mukherjee_electromagnetic_1975}.

The dispersion function Eq.~\eqref{eq:disp_x_lab} can then be plugged in the RTEs Eq.~\eqref{eq:traj_RTEs} to compute the ray trajectory of an X mode in an arbitrarily moving magnetized plasma.  Following this route and examining more specifically the lab-frame group velocity, it was recently shown that the X mode is dragged in the direction of the plasma motion, and that this drag may be significant near resonances and cutoffs, and also for wave frequencies below the lower hybrid frequency~\cite{Langlois2024}. Note however that a condition for Eq.~\eqref{eq:disp_x_lab} to hold is that $\vec k'\perp\vec B_0'$, which may not be verified all along a trajectory for any velocity field.

\section{Conclusion}\label{sec:4_conclusion}

To conclude, after recalling some basic elements of classical geometrical optics, we showed how this method can be used to study the propagation of waves in arbitrarily moving media, in the limit that the velocity field varies slowly in space and time compared to the wavelength and the period of the wave considered. Under this assumption, the lab-frame dispersion function that is required in the trajectory RTEs is obtained by Lorentz transforming the dispersion function assumed to be known in the rest-frame. The trajectory RTEs can then be used to compute the trajectory of the wave similarly to a point-particle.

The trajectory RTEs give in particular access to the group velocity in the lab-frame. We find that, as a result of motion, an additional term depending on $\vec\beta$ appears in the lab-frame group velocity. This term captures the drag of the wave induced by the moving medium. Applying these results to a moving magnetized cold plasma for perpendicular propagation, one finds that the O mode is unaffected by the motion, while the X mode undergoes a drag in the direction of the velocity.

Looking ahead, the work presented here focuses on the ray trajectory and thus only considered the zeroth-order eikonal expansion. As such it cannot inform on the polarization and amplitude evolution. Capturing the effect of motion on polarization, such as polarization drag~\cite{jones_rotary_1976,player_dragging_1976}, requires extending this work to obtain first-order RTEs. This development is currently underway.

\section*{Declaration of interests}

The authors do not work for, advise, own shares in, or receive funds from any organisation that could benefit from this article, and have declared no affiliation other than their research organisations.

\section*{Acknowledgments}

This work was supported by the French Agence Nationale de la Recherche (ANR), under grant ANR-21-CE30-0002 (project WaRP). It has been carried out within the framework of the EUROfusion Consortium, funded by
the European Union via the Euratom Research and Training Programme. The views and opinions expressed herein do not necessarily reflect those of the European Commission. JL acknowledges the support of ENS Paris-Saclay through its Doctoral Grant Program. The authors would like to thank Adrien Bourgoin for valuable discussions.




\def\bysame{\leavevmode ---------\thinspace}
\makeatletter\if@francais\providecommand{\og}{<<~}\providecommand{\fg}{~>>}
\else\gdef\og{``}\gdef\fg{''}\fi\makeatother
\def\cdrandname{\&}
\providecommand\cdrnumero{no.~}
\providecommand{\cdredsname}{eds.}
\providecommand{\cdredname}{ed.}
\providecommand{\cdrchapname}{chap.}
\providecommand{\cdrmastersthesisname}{Memoir}
\providecommand{\cdrphdthesisname}{PhD Thesis}

\end{document}